\documentclass[a4paper,aps, twocolumn,superscriptaddress,nobibnotes,showpacs]{revtex4}
\usepackage[utf8x]{inputenc}
\usepackage[english]{babel}
\usepackage{fontenc}
\usepackage{graphicx}
\usepackage{color}

\begin{document}

\title{Precision Mass Measurements of $^{129-131}$Cd and Their Impact on Stellar Nucleosynthesis via the Rapid Neutron Capture Process}

\author{D. Atanasov}
\affiliation{Max-Planck-Institut f\"{u}r Kernphysik, Saupfercheckweg 1, 69117 Heidelberg, Germany}

\author{P. Ascher}
\affiliation{Max-Planck-Institut f\"{u}r Kernphysik, Saupfercheckweg 1, 69117 Heidelberg, Germany}

\author{K. Blaum}
\affiliation{Max-Planck-Institut f\"{u}r Kernphysik, Saupfercheckweg 1, 69117 Heidelberg, Germany}

\author{R. B. Cakirli}
\affiliation{Department of Physics, University of Istanbul, 34134 Istanbul, Turkey}

\author{T. E. Cocolios}
\affiliation{University of Manchester, Manchester M13 9PL, United Kingdom}

\author{S. George}
\affiliation{Max-Planck-Institut f\"{u}r Kernphysik, Saupfercheckweg 1, 69117 Heidelberg, Germany}

\author{F. Herfurth}
\affiliation{GSI Helmholtzzentrum f\"{u}r Schwerionenforschung GmbH, 64291 Darmstadt, Germany}

\author{D. Kisler}
\affiliation{Max-Planck-Institut f\"{u}r Kernphysik, Saupfercheckweg 1, 69117 Heidelberg, Germany}
 
\author{M. Kowalska}
\affiliation{CERN, 1211 Geneva, Switzerland}

\author{S. Kreim}
\affiliation{Max-Planck-Institut f\"{u}r Kernphysik, Saupfercheckweg 1, 69117 Heidelberg, Germany}
\affiliation{CERN, 1211 Geneva, Switzerland}

\author{Yu. A. Litvinov}
\affiliation{GSI Helmholtzzentrum f\"{u}r Schwerionenforschung GmbH, 64291 Darmstadt, Germany}

\author{D. Lunney}
\affiliation{CSNSM-IN2P3-CNRS, Universit\'{e} Paris-Sud, 91406 Orsay, France}

\author{V. Manea}
\affiliation{CSNSM-IN2P3-CNRS, Universit\'{e} Paris-Sud, 91406 Orsay, France}

\author{D. Neidherr}
\affiliation{GSI Helmholtzzentrum f\"{u}r Schwerionenforschung GmbH, 64291 Darmstadt, Germany}

\author{M. Rosenbusch}
 \affiliation{Ernst-Moritz-Arndt-Universit\"{a}t, Institut f\"{u}r Physik, 17487 Greifswald, Germany}
 
\author{L. Schweikhard}
 \affiliation{Ernst-Moritz-Arndt-Universit\"{a}t, Institut f\"{u}r Physik, 17487 Greifswald, Germany}

\author{A. Welker}
\affiliation{Technische Universit\"{a}t Dresden, 01069 Dresden, Germany}

\author{F. Wienholtz}
 \affiliation{Ernst-Moritz-Arndt-Universit\"{a}t, Institut f\"{u}r Physik, 17487 Greifswald, Germany}
 
\author{R. N. Wolf}
\affiliation{Max-Planck-Institut f\"{u}r Kernphysik, Saupfercheckweg 1, 69117 Heidelberg, Germany}

\author{K. Zuber}
\affiliation{Technische Universit\"{a}t Dresden, 01069 Dresden, Germany}

\date{December 02, 2015}

\begin{abstract}
Masses adjacent to the classical waiting-point nuclide $^{130}$Cd have been measured by using the Penning-
trap spectrometer ISOLTRAP at ISOLDE/CERN. We find a significant deviation of over 400 keV from
earlier values evaluated by using nuclear beta-decay data. The new measurements show the reduction of the
$N = 82$ shell gap below the doubly magic $^{132}$Sn. The nucleosynthesis associated with the ejected wind
from type-II supernovae as well as from compact object binary mergers is studied, by using state-of-the-art
hydrodynamic simulations. We find a consistent and direct impact of the newly measured masses on the
calculated abundances in the $A = 128 - 132$ region and a reduction of the uncertainties from the precision
mass input data.
\end{abstract}


\maketitle

The origin of the elements heavier than iron remains
one of the major quests of today’s observational, experimental, 
and theoretical physics. Produced by neutron
capture reactions \cite{B2FH-RevModPhys.29.547} various isotopes are created in 
radically different environments with time scales ranging from
millions of years \cite{Neyskens-Nature.517.174} 
for the slow ($s$) neutron capture process
to seconds for the rapid ($r$) neutron capture process \cite{Arnould-PhysRep.450.97}. 
Imprints from the nuclear structure are found in the form of
peaks on the solar abundance curve associated with the
closed nuclear shells at neutron magic numbers $N =$ 50, 82 and 126. 
The description of these peaks in astrophysical
simulations naturally has a strong sensitivity to the under-
lying nuclear structure as well as the fundamental choice of
the yet-unknown associated astrophysical scenario.

Presently, the favored sites for the $r$ process are core-collapse 
supernovae and the coalescence of two neutron
stars (NS-NS) or a black hole and a neutron star (BH-NS) \cite{Arnould-PhysRep.450.97}. 
The heat associated with beta-decaying isotopes following an $r$ 
process in the merger scenarios should lead to
observable thermal emission, called a macro- or kilonova \cite{Li-AstrophysJ.507.L59, Kulkarni-arxiv.0510256}. 
The recent observation of an optical transient following 
a gamma-ray burst has been interpreted as such \cite{Tanvir-Nature.500.547, Berger-AstrophysJLett.774.L23}. 
This tantalizing evidence for an $r$-process site has triggered
intense theoretical modeling. A comprehensive study was
recently published \cite{Just-MonthNotRoyAstronSoc.448.541} 
that addresses nucleosynthesis via the $r$ process resulting from NS-NS and BH-NS mergers within
a state-of-the-art hydrodynamic simulation.

However, nuclear data serving as crucial input for the
astrophysical models are still lacking due to the difficulties
in the production and measurement of the required exotic
isotopes. Indeed, many of the nuclides involved lie so far
from stability that they may never be produced in the
laboratory. In this case, nuclear theory is indispensable and
many approaches have been proposed (see \cite{Lunney.RevModPhys.75.1021} for a review).
Whether phenomenological or microscopic, mass models
rely on measured masses for adjusting their parameters.
While measuring masses farther from stability should help
constrain the predictions, the final impact depends on how
many new masses are used, how far they are from what is
known, and the less quantifiable inherent uncertainty of
the model. These points have been addressed in a recent
study \cite{McDonnell.PhysRevLett.114.122501}. Despite regular progress of microscopic nuclear
theory, such as Hartree-Fock-Bogoliubov models and
density functional theory that provide complete and 
consistent data libraries, there are still significant deviations
of predictions from experiment. Therefore, considerable
efforts have been devoted to improve the production yields
and selectivity of exotic nuclear species, as well as the
sensitivity of experimental mass spectrometry.

In this Letter, we report the precision mass measurement
of the closed shell nuclide $^{130}$Cd, previously investigated by
beta-gamma-decay spectroscopy \cite{Dillmann-PhysRevLett.91.162503}, as well as the first
mass determinations of its neighboring isotopes, allowing
further examination of the strength of the $N = 82$ shell
closure beyond the doubly magic $^{132}$Sn. In addition to the
inherent interest in doubly magic nuclides, the high 
abundances of isotopes around magic numbers make their
associated nucleosynthesis sensitive to nuclear physics
input, particularly the $A = 130$ region, as shown by sys-
tematic studies \cite{Mumpower.JPhysG.42.034027}.

The new mass measurements were performed at the on-line 
radioactive ion beam facility ISOLDE/CERN \cite{Kugler-HypInt.129.23} 
using the ISOLTRAP mass spectrometer. The ISOLTRAP
setup consists of a linear segmented radio-frequency quadrupole 
trap (RFQ), a multireflection time-of-flight mass
separator (MR-TOF MS), a preparation, and a precision
Penning trap, each of the latter two placed in the center of a
superconducting magnet \cite{Mukherjee-EurPhysJA, Wolf-NuclInstrumMethodsA.686.82, Kreim-NuclInstrumMethodsB.317.492}.
Depending on the half-life and production yield of the ion of interest, the mass
determination is performed either by the time-of-flight 
ion-cyclotron resonance technique (TOF-ICR) using the 
precision Penning trap \cite{Blaum-PhysRep.425.1} or by performing the time-of-flight
mass spectrometry with the MR-TOF MS \cite{Wienholtz-Nature.498.346}. 

Over the past three decades, TOF-ICR has proven to be
the method of choice in the context of precision mass
measurements of short-lived isotopes \cite{Blaum-PhysScr.2013.014017}. The method is
based on the precise measurement of the cyclotron frequency 
[$\nu_c = q{ B}/(2\pi m)$] of an ion with mass
$m$ and charge $q$ confined in a magnetic field with strength
$B$. The calibration of $B$ is performed before and after a
measurement of the isotope of interest via the cyclotron
frequency $\nu_{c,ref}$ of a reference isotope with a well-known
mass. The frequency ratio $r = \nu_{c,ref}/\nu_{c}$ then yields directly
the mass ratio and allows determining the mass of the ion of
interest \cite{Koenig-IntJMassSpectrom.142.95}. 

The MR-TOF MS recently implemented at ISOLTRAP \cite{Wolf-NuclInstrumMethodsA.686.82, Wolf-IntJMassSpectrom.349-350.123} 
relies on the determination of the ions’ 
flight time ($t$) after multiple revolutions between two
electrostatic mirrors. The time of flight of an ion with
mass $m$ is given by $t = \alpha(m/q)^{1/2} + \beta$, where the two
parameters $\alpha$ and $\beta$ are determined by the flight times $t_{1,2}$ 
of reference isotopes with well-known masses $m_{1,2}$, respectively. 
Substituting $\alpha$ and $\beta$ in the previous formula leads to
a more general relation for the mass of interest:
$$ \sqrt{m} = C_{ToF}(\sqrt{m_1} -\sqrt{m_2}) + 0.5*(\sqrt{m_1} + \sqrt{m_2},$$
where $C_{ToF} = (2t - t_1 -t_2)/2(t_1 - t_2)$ is the experimental 
time-of-flight ratio \cite{Wolf-IntJMassSpectrom.349-350.123}.

The $^{129-131}$Cd isotopes were produced by neutron-induced 
fission in a $50 g/cm^2$ uranium-carbide target. The neutrons were created by a pulsed proton beam with
an energy of 1.4 GeV impinging on a tungsten rod \cite{Koester-EPJA.15.255}. 
The resulting cadmium atoms diffused out of the heated target
through a transfer line to an ion source. A quartz insert in
the transfer line reduced the abundantly produced cesium
and indium contamination \cite{Bouquerel-EurPhysJSpecialTopics.150.277}. 
Element-selective, stepwise resonant photoionization was performed by 
tunable laser radiation \cite{Marsh-RevSciInstrum.85.2}. 

The cadmium ion beam was then
transported towards ISOLTRAP at a kinetic energy of
30 keV through the two-stage high-resolution mass separator. 
The beam entering ISOLTRAP was accumulated in
the RFQ, where it was bunched and cooled via collisions
with helium buffer gas for 20 ms. The ion bunch was then
extracted, and prior to injection in the MR-TOF MS its
energy was adjusted by a pulsed drift cavity to the beam
line potential. Ions were then captured in the MR-TOF MS
by use of the in-trap lift technique \cite{Wolf-IntJMassSpectrom.313.8}. 
After a trapping time for sufficient separation between the ions of interest
and the remaining contaminants, the ions were transported
either to the Penning traps or to an ion detector.
\begin{figure}[b]
\includegraphics[scale=0.4]{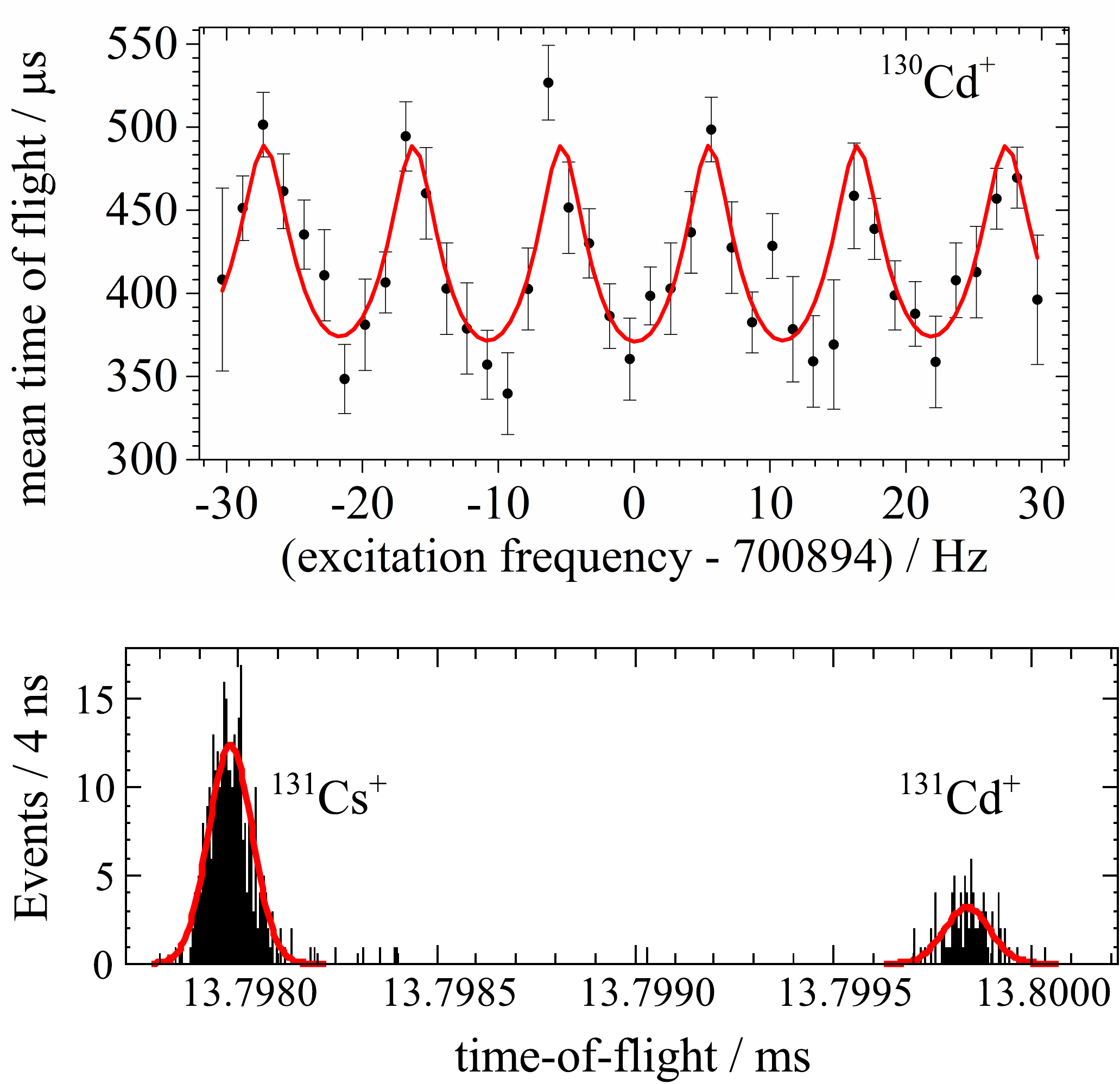}
\caption{(Color online){\it Upper panel}: A typical spectrum showing the TOF-ICR resonance of $^{130}$Cd$^{+}$ ions 
using a Ramsey-type excitation scheme \cite{George-IntJMassSpectrom.264.110}. The line 
represents a fit to the data points where the center frequency corresponds to the 
cyclotron frequency. 
{\it Lower panel}: MR-TOF mass spectrum, i.e. Number of events
as a function of the flight times of $^{131}$Cs$^+$ and $^{131}$Cd$^+$.}
\label{fig-spectra}
\end{figure}

\begin{table*}[t]
 \caption{Frequency ratios ($r = \nu_{c,ref} / \nu_{c}$), time-of-flight ratio ($C_{ToF}$), 
mass excess ($ME$) and the neutron separation energy ($S_n$) of the cadmium isotopes 
measured in this work. Values of the mass excess from the Atomic Mass Evaluation 2012 
(AME12) \cite{Audi-ChinesePhysC.36.1287} are given as well (\# indicates extrapolated values). 
The masses of the references ions used in the evaluation 
are $m(^{131}\rm{Cs}) = 130905465(5) \mu$u and $m(^{133}\rm{Cs}) = 132905451.961(9) \mu$u (from AME12). 
Experimental half-lives are taken from \cite{Lorusso-PhysRevLett.114.192501,Taprogge-PhysRevC.91.054324,Kratz-EurPhysJA.25s01.633}. 
The yield values are only estimates, given the imprecise knowledge of the ISOLTRAP efficiency.}
\begin{ruledtabular}
  \begin{center}
    \begin{tabular}[c]{l c c c c c c c }
          &                     &                 &                 &                            & \multicolumn{2}{c}{\underline{Mass Excess (keV)}}    &       \\ 
     A    & Yield (Ions/$\mu$C) & Half-life (ms)  & Reference       & Ratio $r$ or $C_{ToF}$     & New               & AME12                            & $S_n$ (MeV)      \\ \hline
     129  & 1200                & 151(15),146(8)\footnote{The values correspond to 11/2$^-$ and 3/2$^+$ states, respectively.}   & $^{133}$Cs   & r = 0.970105338(136)  & -63 148(74)\footnote{The mass excess shown in the table is an estimate of the ground-state value based on the measured value of -63 058(17) keV (from the frequency ratio) and the allowance of a ground-state and isomer mixture (see the text for details).}     & -63 510$^{\#}$(200$^{\#}$)    & 3.977(74)  \\
     130  & $>$1000             & 127(2)          & $^{133}$Cs      & r = 0.977645186(180)       & -61 118(22)       & -61 530(160)                     & 6.131(29)  \\
     131  & $>$100              & 98.0(2)         & $^{131,133}$Cs  & $C_{ToF}$ = 0.4823044(539) & -55 215(100)      & -55 331$^{\#}$(196$^{\#}$)       & 2.169(103)
    \end{tabular}
   \end{center} 
   \label{ResTable}
  \end{ruledtabular}
\end{table*}

For the cases of $^{129-131}$Cd$^{+}$, he MR-TOF MS was
employed to provide purified samples for the Penning
traps with trapping times of 1.37 and 13.71 ms, respectively. 
In the preparation Penning trap, the ions were cooled
and recentered for 80 ms in a helium buffer-gas environment. 
Afterwards, the ion bunch was transported to the
precision Penning trap for the TOF-ICR measurement with
a Ramsey-type excitation \cite{George-IntJMassSpectrom.264.110}. 
The excitation timing patterns $\tau^{on}_{rf} - \tau^{off}_{rf} - \tau^{on}_{rf}$ 
used in this experiment were 20-160-20 ms (for $A = 129$) 
and 10-80-10 ms (for $A = 130$). An example of a TOF-ICR resonance curve is presented in 
the upper panel of Fig. \ref{fig-spectra}. It shows the ions’
mean time of flight as a function of the frequency of the quadrupolar
excitation, where the center frequency corresponds to the
cyclotron frequency $\nu_c$ of $^{130}$Cd$^{+}$. In summary, four 
TOF-ICR measurements of $^{129}$Cd were performed, as well as
three of $^{130}$Cd, totaling more than 1500 and 550 events,
respectively.

Considering the low efficiency and the short half-life of $^{131}$Cd$^{+}$, 
the mass measurements were performed by using
the faster MR-TOF MS technique. The calibration of the
device was performed by using off-line reference ions of
stable $^{133}$Cd$^{+}$, as well as the on-line radioactive ions of
surface-ionized $^{131}$Cd$^{+}$ delivered with 
$^{131}$Cd$^{+}$ \cite{Wienholtz-Nature.498.346,Rosenbusch-PhysRevLett.114.202501}. 
In total, 11 spectra were recorded at different numbers of
revolutions 500, 800, and 1000, corresponding to MR-TOF
trapping times of 13.77, 22.02, and 27.51 ms, respectively,
and totaling more than 1350 events. An example of the
obtained spectra is shown in the lower panel of Fig. \ref{fig-spectra}; the
fit method assumes a Gaussian distribution. Our results
are in agreement within statistical uncertainties with the
method using a hybrid Gaussian distribution \cite{Schury-NuclInstrumMethodsB.335.39} for the
peak fits. The final frequency and time-of-flight ratios are
listed in Table \ref{ResTable}.

The beam of $^{129}$Cd$^{+}$ likely contained two nuclear states,
previously identified and measured at ISOLDE. The spins
of the two states were assigned to 11/2 and 3/2 by
hyperfine structure measurements \cite{Yordanov-PhysRevLett.110.192501}. 
The estimated energy difference between the ground and the isomeric
state was inferred from systematics in the cadmium chain
to be about 180(100) keV \cite{Audi-ChinesePhysC.36.1157}. As a definite assignment
of the determined $^{129}$Cd$^{+}$ frequency ratio to one of the
two nuclear states is not possible, an estimation for
the pure ground-state mass excess can be determined
according  to  Appendix  B  of  Ref. \cite{Audi-ChinesePhysC.36.1287}, 
resulting in ME = −63 148(74) keV. 
\begin{figure}[b]
 \includegraphics[scale=0.9]{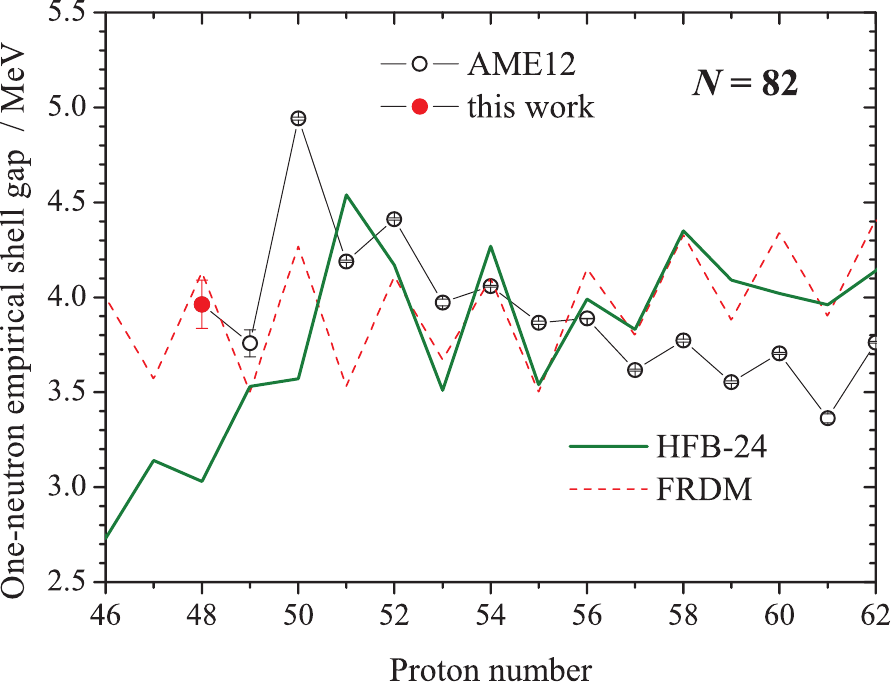}
\caption{(Color online). The empirical one-neutron shell gap. The black open 
circles use the available data from the Atomic Mass Evaluation \cite{Audi-ChinesePhysC.36.1287} 
and the red filled circle is the $^{130}$Cd value calculated using the masses 
from this work. Theoretical values from two mass models are presented for comparison, 
the finite-range droplet model (FRDM) \cite{Moller-AtDataNuclDataTables.59.185} and HFB-24 \cite{Goriely-PhysRevC.88.024308}.}
\label{fig-shell-gap}
\end{figure}
The neutron-separation energies ($S_n$) around the magic
neutron number $N = 82$ presented in Table \ref{ResTable} were computed 
by using the newly determined masses. The drop of $S_n$ 
at the crossing of neutron magic numbers is one of the
important signatures for nuclear magicity, associated with 
large gaps in the spectra of single-particle energies obtained
from shell-model or mean-field approaches.

In agreement with the indications of the earlier beta decay results, 
our precision mass measurements strengthen
and quantify the decrease of the shell strength below $^{132}$Sn. 
Specifically, we observe a reduction of the empirical one-neutron 
shell gap by 1~MeV between $^{132}$Sn ($Z = 50$) and $^{130}$Cd ($Z = 48$), 
also highlighting the doubly magic character of $^{132}$Sn.

In Fig. \ref{fig-shell-gap}, the experimental values of the empirical shell
gap, defined as $S_n(N = 82) - S_n(N = 83)$, are presented. 
Shown in the figure as well are the predictions of two
different mass models, commonly employed to provide
input mass data for various $r$-process calculations. We note
that the microscopic HFB-24 model \cite{Goriely-PhysRevC.88.024308} predicts a significant 
reduction of the empirical shell gap for $Z < 50$, 
while the microscopic-macroscopic finite-range droplet
model \cite{Moller-AtDataNuclDataTables.59.185} predicts a rather constant shell gap, despite the
very close absolute value. 

The $^{130}$Cd s expected to be the progenitor feeding
through $\beta$ decay the second large abundance peak at 
$A \approx 130$ in the Solar System abundance, corresponding to a
region around the stable $^{130}$Te. The impact of mass predictions on the
$r$-process nucleosynthesis in general remains
difficult to ascertain, in the sense that their influence strongly
depends on the adopted astrophysical scenario and most
particularly on the temperature at which the $r$ process takes
place \cite{Arcones-PhysRevC.83.045809}. In the so-called cold $r$ process, 
photodisintegration rates are slow, and consequently no 
$(n, \gamma)-(\gamma,n)$ equilibrium can be achieved. Here nuclear masses influence
the calculated abundances not only through the competition
between the two inverse $(n, \gamma)$ and $(\gamma,n)$ processes, but also
through the neutron capture competition with the $\beta$ decay \cite{Arnould-PhysRep.450.97}.
\begin{figure}[t]
\includegraphics[scale=0.28]{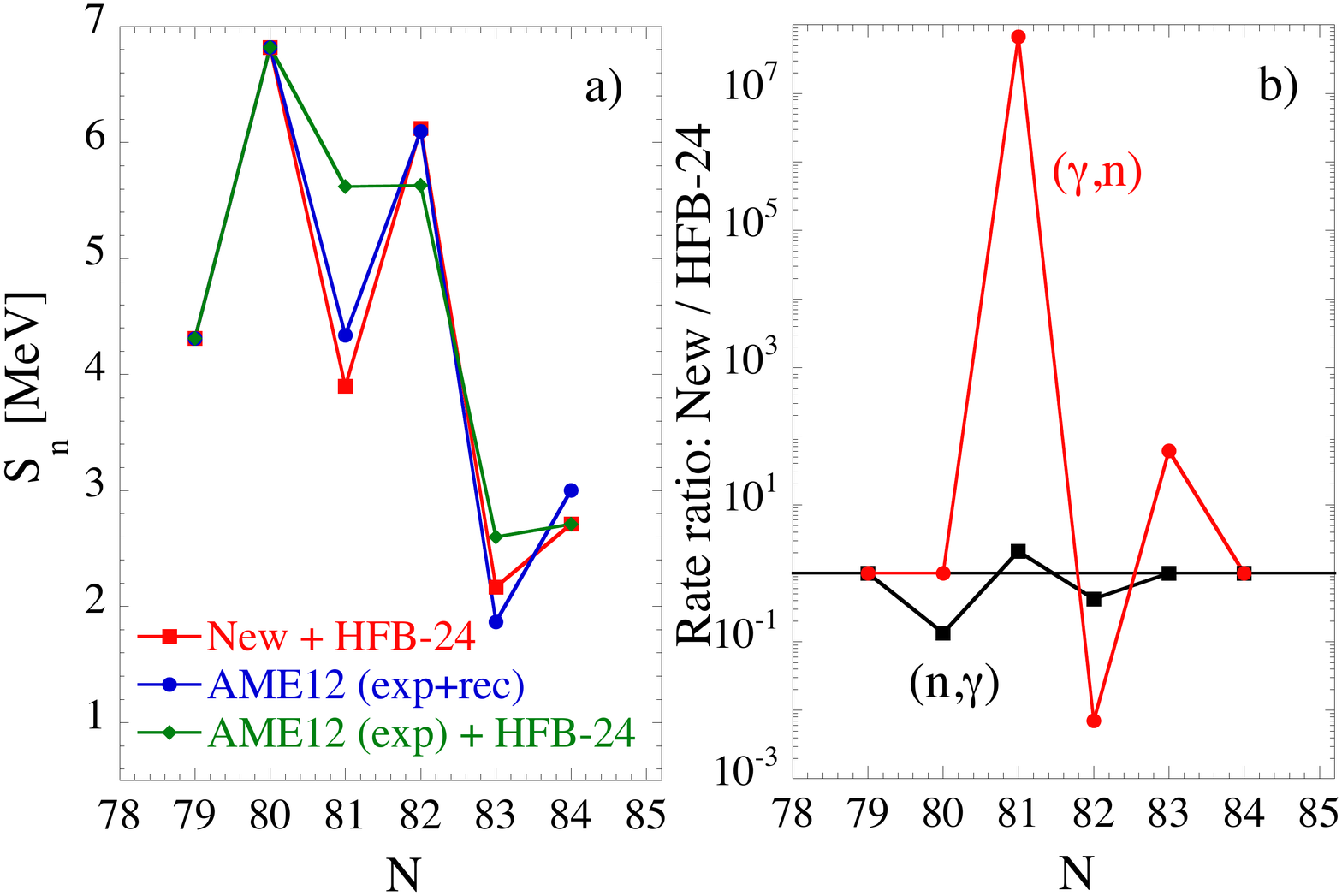}
\caption{ (a) Neutron-separation energies ($S_n$) as a function of neutron number 
 for the newly measured masses (red) in comparison to data in this region from AME12, complemented 
 with extrapolated (recommended) values (blue) or with HFB-24 calculations (green) \cite{Goriely-PhysRevC.88.024308}. 
 (b) Ratio of the neutron capture (n,$\gamma$) and photo-disintegration ($\gamma$,n) rates obtained with 
 experimental masses from this work and the HFB-24 model.}
\label{fig-sn-rate}
\end{figure}

In the present application, the newly measured masses
are used to estimate the neutron capture and 
photodisintegration rates but not the $\beta$-decay half-lives (experimental
half-lives being available in this mass region \cite{Audi-ChinesePhysC.36.1157, Lorusso-PhysRevLett.114.192501}).
The reaction rates are calculated by using the TALYS reaction 
code \cite{Koning-NuclDataScienceandTechnology.211, Goriely-AstronAstrophys.487.767}. 
The impact of the new masses on the 
reaction rates is illustrated in Fig.~\ref{fig-sn-rate} (right panel), where
the Maxwellian-averaged radiative neutron capture and the
photoneutron emission rates at $T = 10^9$~K re compared
when considering a set of nuclear masses from the AME12,
complemented with the masses measured in this work,
or calculations from the HFB-24 model. Since the neutron-
separation energies are affected, the ratio of the neutron
capture to the photoneutron rates is also affected by the new
measurements, so that, even if a $(n, \gamma)-(\gamma,n)$ equilibrium is
established within the Cd isotopic chain, the relative isotopic
abundances may be affected. 

Despite a growing wealth of observational data (see, e.g., Refs.~\cite{Sneden-AnnuRevAstronAstrophys.46.241, Roederer-AstrophysJLett.732.L17}) 
and increasingly better $r$-process models with new astrophysical or 
nuclear physics ingredients, the stellar production site(s) of
$r$-process material has (have) not been identified yet (for a review, see \cite{Arnould-PhysRep.450.97}). 
All proposed scenarios face serious problems. For illustrative purposes,
we consider two widely discussed $r$-process models,
namely, the $\nu$-driven wind model in core-collapse super-
nova explosions of massive stars \cite{Takahashi-AstronAstrophys.286.857, Takahashi-WorldScientific.213, Wanajo-AstrophysJLett.726.L15} 
and the decompression of NS matter during NS-NS and NS-BH mergers,
including the neutrino and viscously driven outflows
generated during the postmerger remnant evolution of
the relic BH-torus system \cite{Goriely-AstrophysJLett.738.L32, Bauswein-AstrophysJ.773.78, Just-MonthNotRoyAstronSoc.448.541}. 
Details concerning the postprocessing of the simulations can be found in
Refs. \cite{Goriely-AstrophysJLett.738.L32, Goriely-PhysRevLett.111.242502}. 

 \begin{figure}
 \includegraphics[scale=0.28]{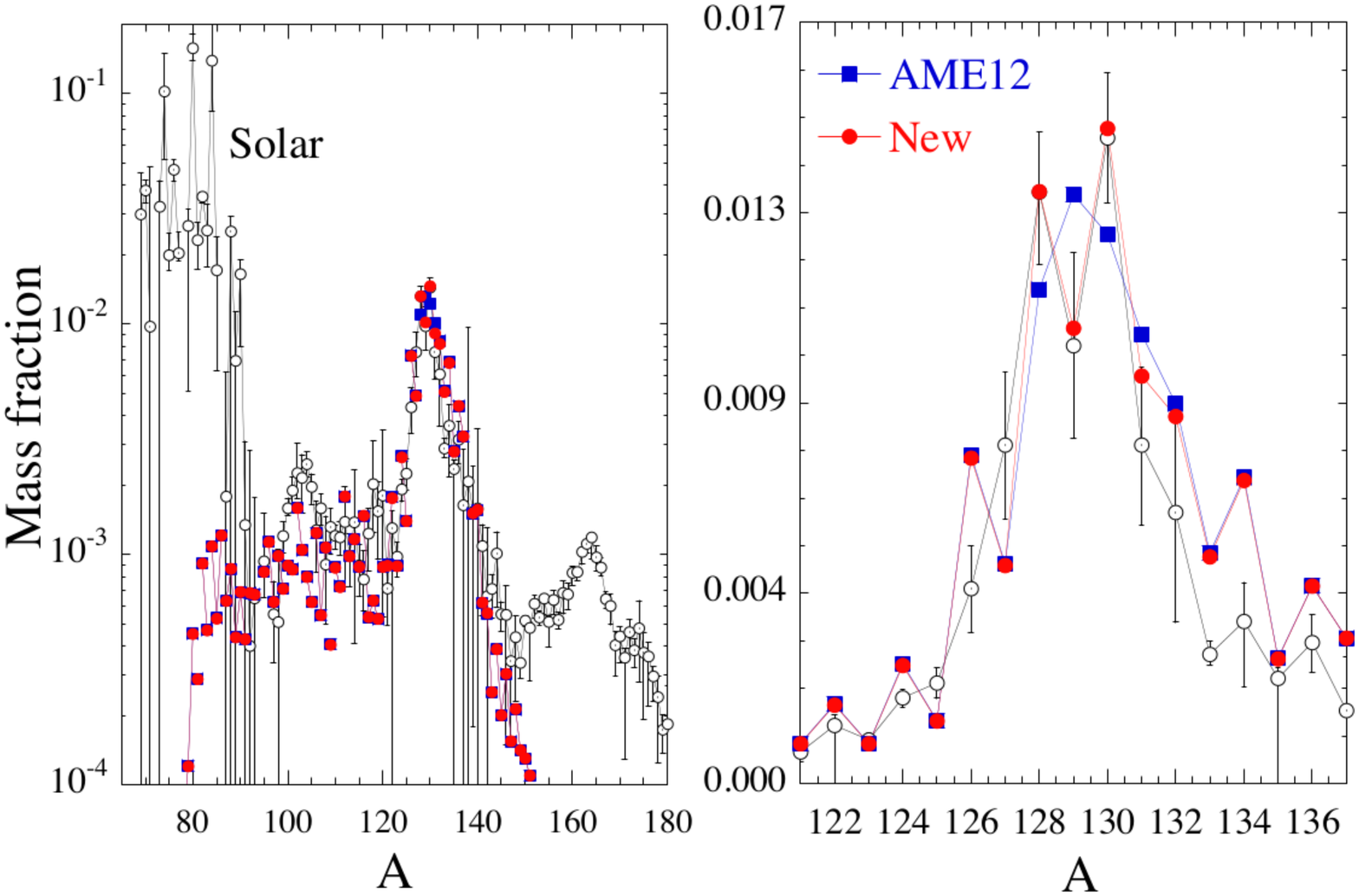}
 \caption{(Color online) Calculated distributions of the $r$-process abundance pattern obtained within the 
 $\nu$-driven wind scenario, 
 see text for conditions and \cite{Arnould-PhysRep.450.97,Takahashi-WorldScientific.213} for more details.
 The blue squares are obtained from AME12 (complemented with HFB-24 masses for 
 experimentally unknown isotopes) and corresponding rates, while the red 
 circles include the new Cd masses. For comparison, the $r$-process solar abundance 
 distribution is shown by open circles. Both theoretical distributions are normalized by the same factor, 
 such that the mass fraction of $^{128}$Te obtained with the new Cd masses reproduces the solar value.}
 \label{fig-apro}
 \end{figure}

In the $\nu$-driven wind scenario, the adopted wind model
corresponds to a subsonic breeze expansion with an
entropy $s_{k_{B}} = 193$, electron fraction $Y_{e} = 0.48$, 
mass loss rate $dM/dt = 6 \times 10^{-7} M_{\odot}$~s$^{-1}$ 
and breeze solution $f_{w} = 3$ 
(see \cite{Arnould-PhysRep.450.97,Takahashi-WorldScientific.213} for more details). 
For such conditions, the $A \simeq 130$ nuclei are dominantly produced and the expansion is
rather fast, so that the neutron irradiation responsible for the
$r$ processing takes place at a rather low temperature and the
final abundance distribution is sensitive to the adopted
neutron capture rates. This specific event is chosen since it
is found to strongly produce isotopes around the second 
$r$-process peak, as shown in Fig.~\ref{fig-apro}. The modified rates based
on the new Cd masses are seen to have an impact in the 
$A \simeq 130$ region. In particular, the odd-even effect between 
$A = 128$ and $A = 130$ is significantly modified due to the 
changes in the neutron-separation energies, especially for 
$^{129}$Cd (Fig.~\ref{fig-sn-rate}). This first example shows that the three
newly measured masses affect directly the
$r$-process abundance distribution in this specific
$\nu$-driven wind scenario, which could potentially explain 
the origin of the Solar System of $r$ nuclei in the vicinity 
of the second $A \simeq 130$ peak despite all the 
remaining uncertainties still
affecting the astrophysical modeling of this site.
  \begin{figure}[t]
  \includegraphics[scale=0.28]{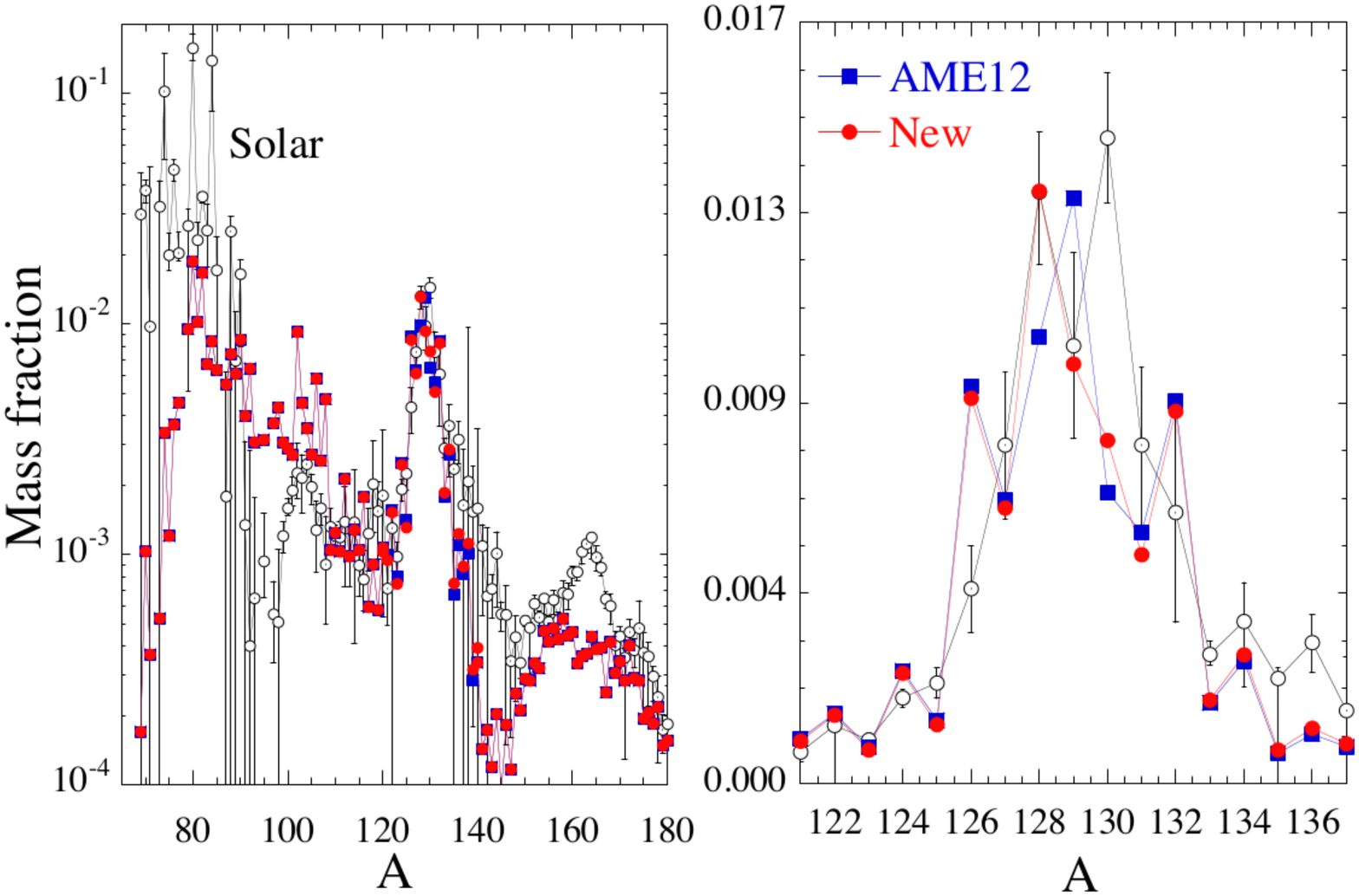}
  \caption{(Color online) Same as Fig. \ref{fig-apro} but showing the final $r$-process abundance curve for 
   viscously driven ejecta from a 3\,$M_\odot$ BH - 0.1\,$M_\odot$ torus system model.}
  \label{fig-nsm}
  \end{figure}
For the compact binary merger scenario, we do not study
the nucleosynthesis in the matter that is dynamically
ejected by tidal and pressure forces during the merging
of the two compact objects but rather in the neutrino and
viscously driven outflows generated during the postmerger
remnant evolution of the relic BH-torus systems. Indeed, in
the prompt ejecta, large neutron-to-seed ratios drive the
nuclear flow into the very heavy-mass region along a
path close to the neutron drip line, leading to multiple
fission recycling at relatively low temperatures, and
essentially $A > 140$ nuclei are found to be produced. In
contrast, the BH-torus ejecta produce heavy elements in the
range from $A \sim 80$ up to thorium and uranium with a
significant contribution to the $A \simeq 130$ abundance peak.
We consider here a representative sample of 310 trajectories 
ejected from a system characterized by a torus mass of 
$0.1M_{odot}$ and a $3M_{odot}$ BH (corresponding to the M3A8m1a5
model of Ref.~\cite{Just-MonthNotRoyAstronSoc.448.541}). The total mass ejected from the BH-torus 
system amounts to $2.5 \times 10^{-2}M_{\odot}$, and the outflow is
characterized by a mean electron fraction $\bar{Y}_{e} = 0.24$, a
mean entropy $\bar{s}/k_{B} = 28$, and a mean velocity 
$\bar{v} = 1.56 \times 10^{9}$~cm/s.

The impact of the newly measured masses is shown in
Fig.~\ref{fig-nsm} and seen to give rise to an abundance peak that is
now shifted by one unit; i.e., the peak location is now at 
$A = 128$ and $A = 129$. Interestingly, despite the fact
that the present distribution results from a mass-weighted
average of hundreds of trajectories, the modification of
only three masses still has an impact on the abundance ratio 
in the corresponding region. This property is linked to the
fact that the masses affect directly the top of the $A = 130$ 
peak. 

In conclusion, the masses of $^{129-131}$Cd were determined
with high precision using the Penning-trap mass spectrometer 
ISOLTRAP. The new masses show a significant reduction of the
$N = 82$ shell gap for $Z < 50$. The new data
provide additional constraints for nuclear theory, considering
the diverging predictions of mass models concerning the 
$N = 82$ empirical shell gap for $Z < 50$. Clearly, the new
measurements bring reliability to the description of
$r$-process 
nucleosynthesis by reducing the uncertainty from the
nuclear-physics input. Given the large volume of data
required for
$r$-process calculations, it is remarkable that only
three masses make an observable impact on the predicted
abundances, highlighting the importance of precision 
measurements in this region of the nuclear chart.

\begin{acknowledgments}
The authors thank G. Audi, CSNSM-IN2P3-CNRS, for
the fruitful discussion in the process of data analysis. We
thank the ISOLDE technical group and the ISOLDE
Collaboration  for  their  support.  We  acknowledge
support by the BMBF (05P12HGCI1, 05P12HGFNE,
05P15HGCIA, and 05P09ODCIA), Nuclear Astrophysics
Virtual Institute (NAVI) of the Helmholtz Association,
Helmholtz-CAS Joint Research Group (HCJRG-108), the
Max-Planck Society, the European Union 7th framework
through ENSAR (Contract No. 262010), the French IN2P3,
the Helmholtz Alliance Program, Contract No. HA216/
EMMI, by the STFC under Grants No. ST/L005743/1 and
No. ST/L005816/1. D. A. acknowledges support by the
IMPRS-PTFS. S. Goriely acknowledges financial support
from FRS-FNRS (Belgium). O. J. acknowledges support
from Max-Planck/Princeton Center for Plasma Physics
(MPPC). S. K. acknowledges support from the Robert-
Bosch Foundation. R. B. C. acknowledges support by the
Max-Planck Partner group.
\end{acknowledgments}

\bibliography{129-131Cd}

\end{document}